\documentclass[aip,jcp,reprint,amssymb,amsmath]{revtex4-1}

\usepackage[T1]{fontenc} 

\usepackage[colorlinks=true]{hyperref} 

\usepackage{xcolor} 

\usepackage{graphicx} 

\begin{document}

\title{The influence of mechanical deformations on surface force measurements}

\author{Romain Lhermerout}
\affiliation{Department of Chemistry, Physical and Theoretical Chemistry Laboratory, University of Oxford, Oxford OX1 3QZ, UK}
\email{romain.lhermerout@chem.ox.ac.uk}

\author{Susan Perkin}
\affiliation{Department of Chemistry, Physical and Theoretical Chemistry Laboratory, University of Oxford, Oxford OX1 3QZ, UK}
\email{susan.perkin@chem.ox.ac.uk}

\begin{abstract}
Experimental investigations of surface forces generally involve two solid bodies of simple and well-defined geometry interacting across a medium. Direct measurement of their surface interaction can be interpreted to reveal fundamental physics in confinement, i.e. independent of the particular geometry. However real solids are deformable -- they can change shape due to their mutual interaction -- and this can influence force measurements. These aspects are frequently not considered, and remain poorly understood. We have performed experiments in a dry atmosphere and across an ionic liquid with a Surface Force Balance (SFB), combining measurement of the surface interactions and simultaneous \textit{in-situ} characterization of the geometry. We show that the mechanical deformations of the surfaces have important consequences for the force measurements, qualitatively and quantitatively. \textcolor{black}{First we find that, whilst the variation of the contact radius with the force across dry nitrogen can be interpreted by the Johnson-Kendall-Roberts (JKR) model, for the (ionic) liquid it is well described only by the Derjaguin-Muller-Toporov (DMT) model}; this contrasts with the previous assertions that SFB experiments are always in the JKR regime. Secondly, we find that mica does not only bend but also experiences a compression. \textcolor{black}{By performing experiments with substantially thicker mica than usual we were able to investigate this with high precision, and find compression of order $1~\mathrm{nm}$ with $7~\mathrm{\mu m}$ mica. These findings imply that, in some cases, (i)~ the procedure to calibrate mica thickness has to be revisited, and (ii)~structural forces measured across nanoconfined liquids must be interpreted as a convolution of the surface forces across the liquid and the mechanical response of the confining solids. We show an example in which the detailed shape of the measured structural force profile cannot be described by the usual exponentially decaying harmonic oscillation, but is well fitted by an heuristic equation supposing that mica compression is dominant over liquid compression.} We discuss the influence of mica thickness, and propose a scaling criterion to distinguish situations where the solid deformation is negligible and when it is dominant.
\end{abstract}

\maketitle


\section{Introduction}
\label{sec:Introduction}

\begin{figure*}
\centering
\includegraphics{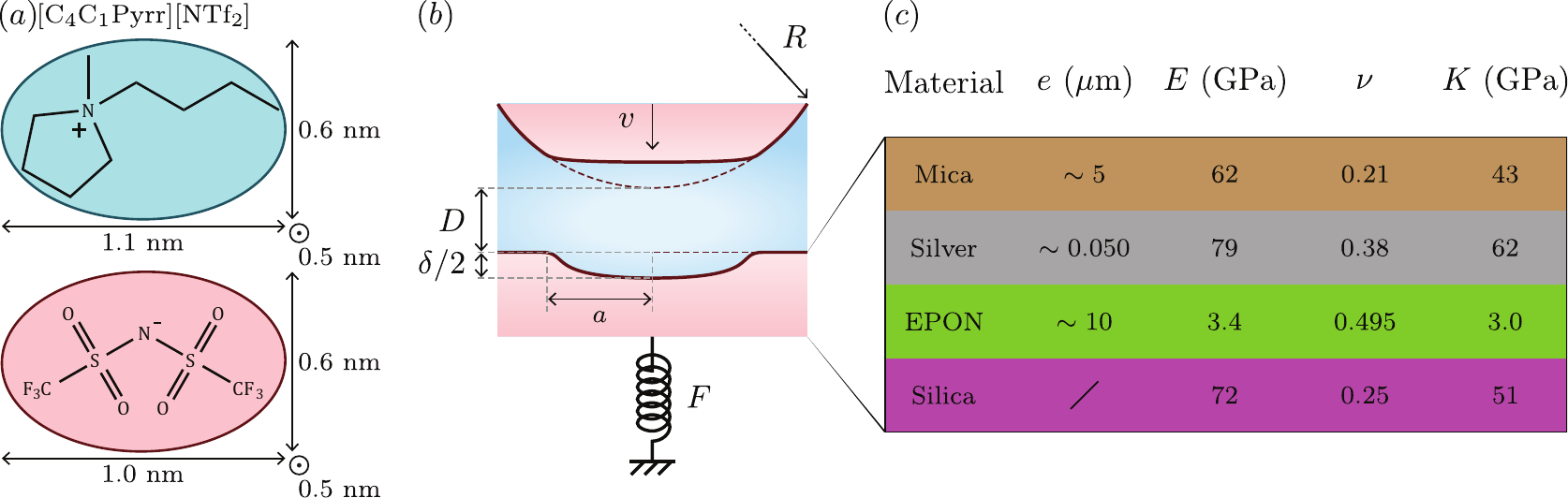}
\caption{(a)~Chemical structure and sizes of [C$_4$C$_1$Pyrr][NTf$_2$]. Ion sizes are estimated from geometry, bond lengths and covalent radii, associated with the most stable configuration found by energy minimisation (Chem3D 16.0, PerkinElmer Informatics). (b)~Schematic of the SFB experiment to measure the surface interactions and to characterize \textit{in-situ} the geometry, when a liquid is confined between two mica surfaces. (c)~Composition of the layers forming each solid surface, with associated thickness~$e$, Young's modulus~$E$, Poisson's ratio~$\nu$ and elastic modulus~$K=\frac{2}{3}\frac{E}{1-\nu^2}$ (values from~\cite{McNeil1993a,McGuiggan2007a,Koster1961a}). \footnote{\textcolor{black}{For the anisotropic mica, the given mechanical properties correspond to the \textit{c}-axis, the direction of interest for this study. A wide range of Young's moduli have been reported for mica ($50-500~\mathrm{GPa}$ in~\cite{Horn1987a}); the quoted value has been consistently obtained by Brillouin scattering~\cite{McNeil1993a} and nanoindentation~\cite{McGuiggan2007a}.}}}
\label{Fig1}
\end{figure*}

Understanding the behavior of liquids in nano-confinement is crucial for a range of applications including energy storage (electrolyte in contact with porous electrodes), lubrication (lubricant in between rough sliding surfaces) and filtration (like sea water through a membrane), as well as in biological systems (ion crossing the cell membrane in a nano-channel, etc.). Very often, model experiments are performed with elementary systems of simple geometry: two surfaces facing each other in force measurements, a single channel in flow measurements. This allows for a simpler mathematical description of the system, while putting aside the complex collective effects happening in the presence of multiple asperities or pores. The ultimate goal is to understand the underlying physics at a fundamental level, in particular independently of a particular geometry. For example, the Surface Force Apparatus/Balance (SFA/SFB), and the Atomic Force Microscope (AFM) are force measurements techniques using surfaces of very different radii of curvatures~$R$: crossed cylinders of radii~$\sim 1~\mathrm{cm}$ in SFA, sphere of radius~$\sim 10~\mathrm{nm} - 10~\mathrm{\mu m}$ and a plane in AFM. The normal interaction force profiles~$F(D)$ between crossed cylinders or sphere and plane can, in certain conditions, be directly compared by rescaling the measured force~$F$ by the radius of curvature~$R$. Indeed, Derjaguin showed that~$F/R$ is proportional to the energy density between equivalent planar and parallel surfaces, provided that (i)~$D\ll R$, (ii)~the interactions involved are additive and (iii)~the surfaces are not deformed~\cite{Israelachvili2011a}. However, real solids are not perfectly rigid, and significant deformations of the surfaces can occur depending on the strength and range of the interaction. For example, when measuring friction in the boundary lubrication regime and under applied load, a strong repulsion across the molecularly-thick boundary film typically leads to a substantial flattening of the surfaces. These mechanical deformations have to be taken into account in order to interpret correctly the data, to compare with theories or numerical simulations, and to extrapolate the results to other experimental set-ups or practical applications. The SFA is a tool of choice in this respect, because the analysis of the so called Fringes of Equal Chromatic Order (FECO) allows for an \textit{in-situ} characterization of the geometry, with a sub-molecular resolution ($\sim 0.1~\mathrm{nm}$) in the normal direction and an optical resolution ($\sim 1~\mathrm{\mu m}$) in the lateral direction. So far, only a few experimental investigations have been dedicated to the study of contact mechanics with the SFA (see for example~\cite{Horn1987a,McGuiggan2007a}), and surface deformations are not measured in a majority of the SFA studies, because they are assumed to have a negligible effect on the measurements or because they are calculated using theoretical model~\cite{Klein1995a,Klein1998a,Kumacheva1998a,Perkin2006a,Klein2007a,Mazuyer2008a,Smith2013a}.

In this paper, we show that it is important to measure the deformations of the surfaces in the SFA, because these mechanical deformations can strongly contribute to the shape of the force profile and are tricky to model quantitatively given the complexity of the system (a surface consisting of a mica sheet glued on a glass lens, the solid is composed of several layers with different thicknesses and elasticities). For this purpose, we performed SFB measurements in dry atmosphere and across an ionic liquid, exploiting the full capabilities of the instrument to measure the surface interactions and to characterize \textit{in-situ} the geometry. In the first case of the dry atmosphere (strong adhesion), we find that the variation of the contact radius with the force is well described by a Johnson-Kendall-Roberts (JKR) model and an effective elastic modulus describing the layered system, in agreement with~\cite{Horn1987a}. Contrary to what is commonly asserted, we observe that the compression of mica is not negligible, and that models have to include the finite size of the mica in order to fit the variation of the mica indentation with the force. As a broad consequence, the common calibration procedure that uses the jump-in point in dry atmosphere leads to an underestimation of the mica thickness and an equivalent outward shift of the force profile measured after injecting the liquid. The error is $\sim 1~\mathrm{nm}$ for a $\sim 7~\mathrm{\mu m}$-thick mica, and is expected to decrease with the mica thickness, albeit always present. In the second case of the ionic liquid (moderate adhesion), we find that the variation of the contact radius with the force is well described by a Derjaguin-Muller-Toporov (DMT) model and an effective elastic modulus describing the layered system, in contrast with the previous assertion that contact mechanics between adhering surfaces in the SFA is always described by JKR model~\cite{Christenson1996a}. We also show that mechanical deformations strongly affect the shape of the structural force profile, with the commonly used exponentially decaying harmonic oscillation being convoluted with the compression of the mica. This effect is more important and independent of the mica thickness at low loads, and smaller and reduced for thinner mica at high loads. We propose a heuristic formulation to describe such convoluted structural force profile when the mica compression dominates the liquid compression, as well as a general criterion to distinguish the two opposite regimes of convolution (i.e. mica compression negligible or dominant compared to liquid compression).

The paper is organized as follows. In section titled ``Models, Materials and Methods'', we first recall the models of contact mechanics that will be needed to analyze the data, then we summarize the general procedure to perform a SFB experiment, there we present the protocol to determine the surface deformations. In section titled ``Results and Discussion'', we describe and interpret the measurements performed in dry atmosphere and across an ionic liquid.

\section{Models, Materials and Methods}
\label{sec:ModelsMaterialsandMethods}

\subsection{Models of contact mechanics}
\label{subsec:Modelsofcontactmechanics}

In the following, we recall the hypotheses and consequences associated to the different models of contact mechanics, that will be used to analyze the measurements (reviews can be found in~\cite{Israelachvili2011a,Horn1987a,Maugis1992a,Grierson2005a}). All the models presented here rely on a common set of hypotheses or approximations. The two solid bodies in contact are supposed (i)~semi-infinite, (ii)~composed of linear (no plasticity) homogeneous (no stack of layers) isotropic (not crystalline) and purely elastic (no viscosity) materials, (iii)~with perfectly smooth and frictionless surfaces, and (iv)~in a regime where the contact radius is much smaller than their radius of curvature: $a \ll R$. These models differ in the way adhesion is taken into account. In the following, we will consider the simple geometry of two cylinders of same radius of curvature~$R$ and material that are crossed at 90 degrees (or equivalently a sphere of radius of curvature~$R$ and a plane of same material), with the top solid controlled in position and the bottom solid mounted on a spring of stiffness~$k$ (see sketch in Figure~\ref{Fig1}(b)).

In Hertz model~\cite{Hertz1882a}, it is hypothesized that there are no attractive forces between the surfaces (hard wall interaction, acting inside the contact area only). The normal force (or load)~$F$, contact radius~$a$ and indentation~$\delta$ (defined positive for compression and negative for dilatation) are related by:

\begin{equation}
\left\{
\begin{array}{lll}
F&=&\frac{K a^3}{R} \\[5pt]
\delta &=&\frac{a^2}{R}
\end{array}
\right.\textrm{~,}
\label{eq:Hertz_F_a_delta_a}
\end{equation}

\noindent where $K=\frac{2}{3}\frac{E}{1-\nu^2}$ is the elastic modulus, with $E$ the Young's modulus and $\nu$ the Poisson's ratio. At lateral scale~$|x| \ll R$, the distance~$z$ between the surfaces is given by:

\begin{equation}
z= \left\vert\,
\begin{array}{l}
\frac{a^2}{\pi R}\left[\sqrt{\left(\frac{x}{a}\right)^2-1}+\left(\left(\frac{x}{a}\right)^2-2\right)\arctan\sqrt{\left(\frac{x}{a}\right)^2-1}\right]\\[10pt]
\qquad\mathrm{for~} |x|\geq a\\
0\\
\qquad\mathrm{for~} |x|<a
\end{array}
\right. \textrm{.}
\label{eq:Hertz_z_x}
\end{equation}

\noindent When moving up the top solid, the surfaces separate at $F_\mathrm{s}=0$, $a_\mathrm{s}=0$, $\delta_\mathrm{s}=0$, with no jump-out.

In Derjaguin-Muller-Toporov (DMT) model~\cite{Derjaguin1975a}, attractive forces of finite range are added (sticky hard wall interaction, acting inside the contact area and in a ring-shaped zone outside the contact area), but adhesion is assumed not to deform the surfaces, leading to a discontinuity of the normal stress at the edge of the contact area. The normal force (or load)~$F$, contact radius~$a$ and indentation~$\delta$ are related by:

\begin{equation}
\left\{
\begin{array}{lll}
F&=&\frac{K a^3}{R}-2\pi RW \\[10pt]
\delta &=&\frac{a^2}{R}
\end{array}
\right.\textrm{~,}
\label{eq:DMT_F_a_delta_a}
\end{equation}

\noindent where $W$ is the adhesion energy (per unit area, taken positive), and the deformation profile~$z(x)$ is the same than in the Hertz model. When moving up the top solid, the surfaces separate at $F_\mathrm{s}=-2\pi RW$, $a_\mathrm{s}=0$, $\delta_\mathrm{s}=0$, with a jump-out over a distance~$2\pi RW/k$ due to the spring instability.

In Johnson-Kendall-Roberts (JKR) model~\cite{Johnson1971a}, attractive forces of zero range are added (Baxter i.e. infinitely short range square interaction, acting only inside the contact area), and adhesion can deform the surfaces, leading to a divergence of the normal stress at the edge of the contact area. The normal force (or load)~$F$, contact radius~$a$ and indentation~$\delta$ are related by:

\begin{equation}
\left\{
\begin{array}{lll}
F&=&\left[\sqrt{\frac{K a^3}{R}}-\sqrt{\frac{3}{2}\pi R W}\right]^2-\frac{3}{2}\pi R W \\[10pt]
\delta &=&\frac{a^2}{R}\left[1-\frac{4}{3}\left(\frac{a_\mathrm{s}}{a}\right)^{3/2}\right]
\end{array}
\right.\textrm{~,}
\label{eq:JKR_F_a_delta_a}
\end{equation}

\noindent with $a_\mathrm{s}=\left(\frac{3\pi R^2 W}{2K}\right)^{1/3}$. At lateral scale~$|x| \ll R$, the distance~$z$ between the surfaces is given by:

\begin{equation}
z= \left\vert\,
\begin{array}{l}
\frac{a^2}{\pi R}\left[\sqrt{\left(\frac{x}{a}\right)^2-1}\right.\\
\left. +\left(\left(\frac{x}{a}\right)^2-2+\frac{8}{3}\left(\frac{a_\mathrm{s}}{a}\right)^{3/2}\right)\arctan\sqrt{\left(\frac{x}{a}\right)^2-1}\right]\\[10pt]
\qquad\mathrm{for~} |x|\geq a\\
0\\
\qquad\mathrm{for~} |x|<a
\end{array}
\right. \textrm{.}
\label{eq:JKR_z_x}
\end{equation}

\noindent When moving up the top solid, the surfaces separate at the point where~$\frac{\mathrm{d}F}{\mathrm{d}\delta}=-k$. If the spring constant is low enough, this condition can be approximated by~$\frac{\mathrm{d}F}{\mathrm{d}\delta}=0$, and the surfaces separate at $F_\mathrm{s}=-\frac{3}{2}\pi RW$, $a_\mathrm{s}=\left(\frac{3\pi R^2 W}{2K}\right)^{1/3}$, $\delta_\mathrm{s}=-\left(\frac{\pi^2 R W^2}{12K^2}\right)^{1/3}$, with a jump-out over a distance~$\frac{3}{2}\pi RW/k$.

In Maugis model~\cite{Maugis1992a}, attractive forces of finite range~$d$ are added (Dugdale i.e. square-well interaction, acting inside the contact area and in a ring-shaped zone outside the contact area), and adhesion can deform the surfaces, leading to a normal stress that presents no singularity at the edge of the contact area. The normal force (or load)~$F$, contact radius~$a$ and indentation~$\delta$ are related by implicit equations, together with the dimensionless parameter~$\mathrm{Ma}$:

\begin{equation}
\mathrm{Ma}=\left(\frac{8RW^2}{\pi K^2 d^3}\right)^{1/3}\textrm{~.}
\label{eq:Maugis_Ma}
\end{equation}

\noindent Physically, $\mathrm{Ma}$ is the ratio between the elastic indentation due to adhesion and the range of the attractive forces themselves. The three previous models are special cases of Maugis model: Hertz limit corresponds to $\mathrm{Ma}=0$, DMT applies for $\mathrm{Ma}\ll 1$, and JKR is recovered for $\mathrm{Ma}\gg 1$. In the transition regime $\mathrm{Ma}\sim 1$, none of the DMT and JKR models are valid and the implicit equations from Maugis model have to be used to describe the contact mechanics.


\subsection{Surface Force Balance}
\label{subsec:SurfaceForceBalance}

The way the Surface Force Balance (SFB) works has been explained in details in previous publications~\cite{Israelachvili2011a,Perkin2006a,Lhermerout2018b}. Here we briefly recall the principle of the instrument, illustrated in Figure~\ref{Fig1}(b), and the details particular to the present experiments.

Muscovite mica is cleaved to produce atomically-smooth facets of micrometric thickness and millimetric extension, that are backsilvered and glued onto glass (fused silica) cylindrical (radius~$R\sim~1\mathrm{cm}$) lenses with an epoxy resin (EPON 1004, Shell Chemicals). Two surfaces are made with mica of the same thickness, form a stack of different layers (represented in Figure~\ref{Fig1}(c), together with their thicknesses, Young's moduli, Poisson's ratios and elastic moduli), and are arranged in a cross-cylinder geometry. First, calibrations are done in a dry atmosphere, which is achieved by inserting P$_2$O$_5$, phosphorus pentoxide (Sigma-Aldrich, 99\%), in the chamber and purging the chamber with N$_2$, nitrogen, during about one hour prior to the calibrations. Secondly, measurements are performed with an ionic liquid, because these liquids have proved to resist squeeze-out even under very large loads~\cite{Lhermerout2018a}, a regime in which significant mechanical deformations of the surfaces are expected. The liquid used is [C$_4$C$_1$Pyrr][NTf$_2$], 1-butyl-1-methylpyrrolidinium \textit{bis}[(trifluoromethane)sulfonyl]imide (Iolitec, 99\%), which chemical structure and sizes are indicated in Figure~\ref{Fig1}(a) (molar mass $M=422.41~\mathrm{g/mol}$, density $\rho=~1.405\mathrm{g/mL}$, refractive index $n=1.422$ and dynamic viscosity $\eta=74~\mathrm{mPa.s}$ at $25^\circ \mathrm{C}$~\cite{Rao2011a}). It is dried in a Schlenck line at $60^\circ \mathrm{C}$ and $5 \times 10^{-3}~\mathrm{mbar}$ for $\sim~10$~hours and inserted in the chamber just after, which contains P$_2$O$_5$ and is purged again with N$_2$ during about one hour prior to the measurements.

White light is passed through the confined medium, interferes in this optical resonator, and is then directed towards a spectrometer and collected by a CCD camera (QImaging Retiga R6, resolution $2688 \times 2200~\mathrm{px}^2$). The analysis of the Fringes of Equal Chromatic Order (FECO, shown in Figures~\ref{Fig2}(a) and~\ref{Fig2}(c)) then allows to deduce the apical distance~$D$~\cite{Israelachvili1973a}, and to characterize \textit{in-situ} the contact geometry (procedure detailed in next subsection). $D$ is measured with a precision of~$0.02~\mathrm{nm}$ given by the standard deviation of the signal, and an accuracy of~$1~\mathrm{nm}$ due to light disalignment when changing the contact spot~\cite{Schwenzfeier2019a}.

The top surface can be moved normally with a stepper motor (large displacement range $\sim 10~\mathrm{\mu m}$, poor linearity, mechanical vibrations induced) or with a piezoelectric tube (small displacement range $\sim 1~\mathrm{\mu m}$, good linearity, no measurable mechanical vibrations induced). For a given run, the velocity~$v$ can be determined with a precision of $\sim 1\%$. From run to run, this velocity can typically vary by $\sim 10\%$ for the same control parameters, because of thermal drifts. In the following, some graphs result from the superposition of several runs, that is why the indicated velocities have to be associated with an error bar of $\sim 10\%$. The bottom lens is mounted on a spring of constant $k=2670 \pm 84~\mathrm{N/m}$, which is calibrated before the experiment by measuring its deflection when adding different masses. The normal force~$F$ is then deduced from the temporal evolution of the distance~$D(t)$ when applying a constant velocity~$v$ to the top surface, using a procedure that takes into account the viscous force that is not negligible at large separations (detailed in~\cite{Lhermerout2018b}). In comparison to previous studies performed in our group, the normal spring is here about 20 times stiffer, in order to apply larger load (at fixed displacement range) and so to induce substantial mechanical deformations of the surfaces of particular interest in the present study. This comes with a price in terms of sensitivity limit, $\sim 10^{-2}~\mathrm{mN}$, which doesn't allow to detect the anomalously long-range electrostatic force that has been observed with concentrated electrolytes~\cite{Gebbie2015a,Smith2016a}. \textcolor{black}{This choice of a stiff spring is technically convenient to study contact mechanics, but does not affect any of the conclusions of this study. Using a softer spring would just reduce the explored range of load, and decrease the slope of the spring instability.}


\subsection{Determination of surface deformations}
\label{subsec:Determinationofsurfacedeformations}

\begin{figure*}
\centering
\includegraphics{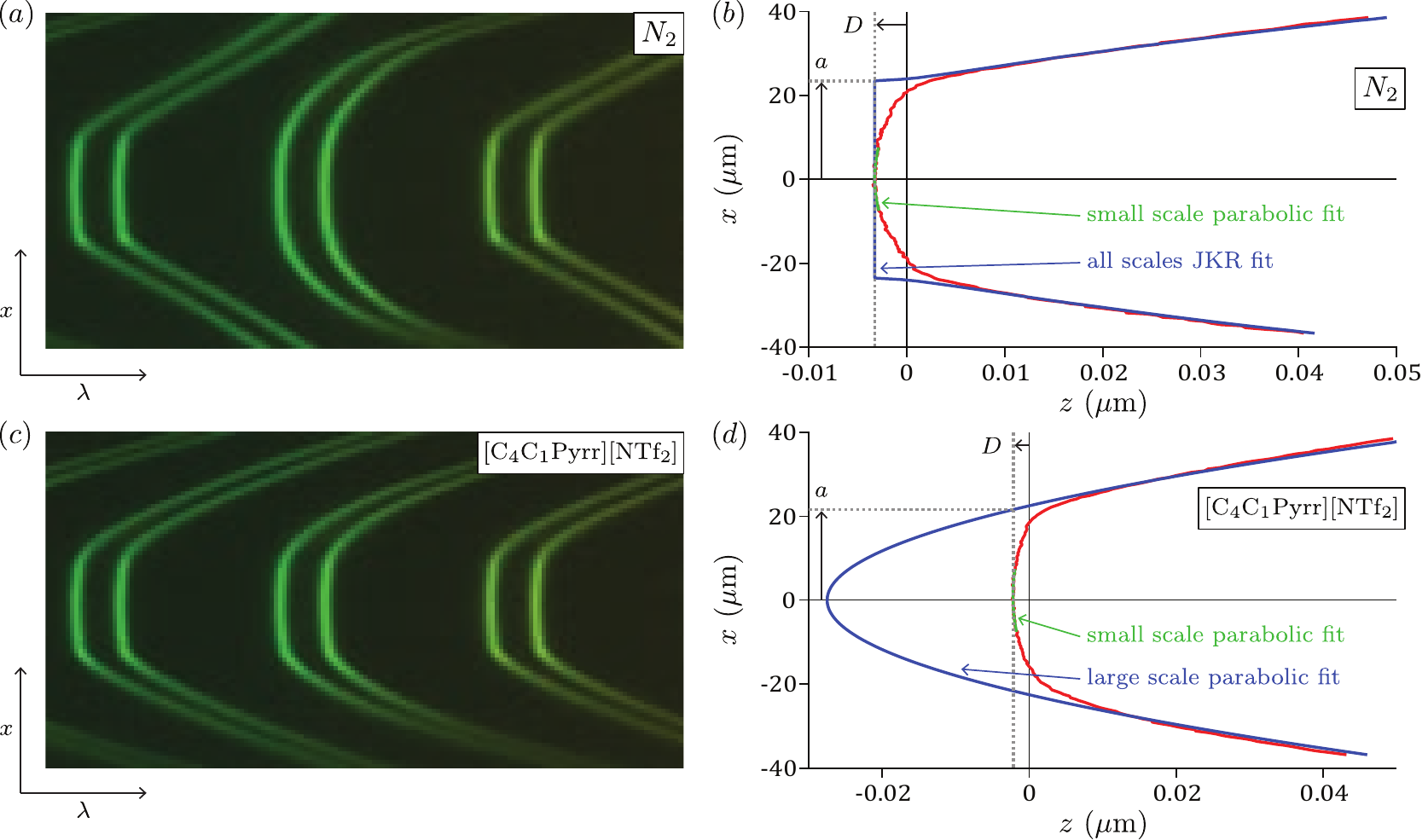}
\caption{(a)~Picture of the FECO when the two solid surfaces are in contact across N$_2$, observed in (wavelength~$\lambda$)-(lateral distance~$x$) space. (b)~Corresponding profile of the distance~$z$ between the surfaces along the lateral coordinate~$x$ (in red). A parabolic fit at small scale close to the apex (in green) allows to measure the apical distance~$D$, while a fit with the JKR profile (equation~\ref{eq:JKR_z_x}) at all measured scales (in blue) is used to extract the contact radius~$a$. (c)~Picture of the FECO when the two solid surfaces are in contact across [C$_4$C$_1$Pyrr][NTf$_2$], observed in (wavelength~$\lambda$)-(lateral distance~$x$) space. (d)~Corresponding profile of the distance~$z$ between the surfaces along the lateral coordinate~$x$ (in red). A parabolic fit at small scale close to the apex (in green) allows to measure the apical distance~$D$, while a parabolic fit at large scale (in blue) is used to extract the contact radius~$a$. In both cases, the FECO images were recorded with a black and white camera, then recolored using the calibration of the wavelength axis with a mercury lamp. The two particular cases shown here correspond to the points of maximum load reached in Figure~\ref{Fig3}.}
\label{Fig2}
\end{figure*}

In this subsection, we explain in details the procedure of analysis of the FECO to deduce the apical distance~$D$ and the geometry of the surfaces, i.e. the radius of curvature~$R$ and the contact radius~$a$ (defined in Figure~\ref{Fig1}(b)). In general, the glue used to prepare the surfaces is heterogeneous in thickness, leading to a local radius of curvature of mica that is different from the radius of curvature of the glass lens below and differs from one surface to the other (by typically~$\sim 10\%$), and to an contact zone of elliptic symmetry when crossing the two cylinders at right angle. During the experiments, we observe the FECO along only one direction~$x$ parallel to the axis of symmetry of one lens (typical FECO images for N$_2$ and [C$_4$C$_1$Pyrr][NTf$_2$] cases shown in Figures~\ref{Fig2}(a) and~\ref{Fig2}(b)), that is why to interpret the data we will suppose that the surface deformation is the same in the perpendicular direction~$y$, i.e. that the contact zone has a circular symmetry.

In (wavelength~$\lambda$)-(lateral distance~$x$) space first, we measure the shape~$\lambda_p (x)$ of fringe of odd order~$p$. For each line, the fringe position is detected by calculating the center of mass of the doublet (due to mica birefringence) after applying a threshold on the image (when the intensity is smaller than the threshold, it is set equal to the threshold). The threshold is chosen just above the intensity fluctuations of the background, to reduce the noise on the signal. Providing that the mica thickness is supposed constant and is known, the separation profile between the mica surfaces~$z(x)$ can then be deduced (typical profiles for N$_2$ and [C$_4$C$_1$Pyrr][NTf$_2$] cases shown in Figures~\ref{Fig2}(b) and~\ref{Fig2}(d))~\cite{Israelachvili1973a}. The separation profile is measured up to a maximum scale~$z_\mathrm{max}\sim 50~\mathrm{nm}\ll R\sim 1~\mathrm{cm}$, that is why the undeformed shape (when the surfaces are far from contact), circular in theory, is here observed locally and very well described by a parabola. There, a fitting procedure is used to extract $D$, $R$ and $a$. On one hand, a parabolic fit is done at small scale close to the apex (green curve in Figure~\ref{Fig2}(b) and Figure~\ref{Fig2}(d)), with 3 free parameters, providing the apical distance~$D$ (negative in the present cases shown in the Figures, as explained in details in the next section). On the other hand, $R$ and $a$ are obtained with different methods, depending whether the solid surfaces are separated by N$_2$ or by [C$_4$C$_1$Pyrr][NTf$_2$].

In the case of [C$_4$C$_1$Pyrr][NTf$_2$] (moderate adhesion), the mechanical deformations are limited to a scale~$z\ll z_\mathrm{max}$, and the separation profile matches the undeformed shape at large measurable distances~$z \sim z_\mathrm{max}$. A second parabolic fit is done at large scale only (blue curve in Figure~\ref{Fig2}(d)), with a function of the form:

\begin{equation}
z=z_0+\frac{\left(x-x_0\right)^2}{2R} \textrm{,}
\label{eq:z_x_parabolic_fit}
\end{equation}

\noindent where $x_0$ and $z_0$ are 2 free parameters controlling the position of the parabola, and $R=0.92 \pm 0.01~\mathrm{cm}$ is the radius of curvature that is adjusted using one image when the surfaces are far from contact and then kept fixed. By definition, the contact radius is the lateral distance~$|x-x_0|$ at which the extrapolated undeformed profile crosses the contact plane at~$z=D$, and is simply given by $a=\sqrt{2R\left(D-z_0\right)}$. To do this parabolic fit at large scale only, the points associated to values $|x-x_0|<a$ are excluded from the fit, and the fitting procedure is repeated iteratively: on first iteration no point is excluded and a value $a_1$ is deduced, on second iteration $a_1$ is used to exclude some points from the fit and $a_2$ is deduced, etc. In practice, 3 iterations are enough for the value of $a$ to converge, as additional iterations lead to insensitive changes.

In the case of N$_2$ (strong adhesion), the mechanical deformations are present at all measurable scales~$z\leq z_\mathrm{max}$, and the undeformed region of the profile cannot be observed. That is why the general definition of the contact radius~$a$ cannot be used, and a model is needed to fit the deformation. A fit of the separation profile is done at all the measurable scales (blue curve in Figure~\ref{Fig2}(b)), with a function derived from the JKR model (equation~\ref{eq:JKR_z_x}):

\begin{widetext}
\begin{equation}
z= \left\vert\,
\begin{array}{l}
D+\frac{a^2}{\pi R}\left[\sqrt{\left(\frac{x-x_0}{a}\right)^2-1}+\left(\left(\frac{x-x_0}{a}\right)^2-2+\frac{8}{3}\left(\frac{a_\mathrm{s}}{a}\right)^{3/2}\right)\arctan\sqrt{\left(\frac{x-x_0}{a}\right)^2-1}\right]\quad\mathrm{for~} |x-x_0|\geq a\\[10pt]
D\quad\mathrm{for~} |x-x_0|<a
\end{array}
\right. \textrm{,}
\label{eq:z_x_JKR_fit}
\end{equation}
\end{widetext}

\noindent where the contact radius~$a$ and the center of the contact zone~$x_0$ are 2 free parameters, the apical distance~$D$ is known from the small scale parabolic fit, the radius of curvature~$R=0.92 \pm 0.01~\mathrm{cm}$ is adjusted using one image when the surfaces are far from contact and then kept fixed, and the contact radius at the jump-out point~$a_\mathrm{s}=11.30~\mathrm{\mu m}$ is adjusted using the image just before the jump-out of the surfaces and then kept fixed. It is clearly visible in Figure~\ref{Fig2}(b) that the JKR model doesn't fit well the deformed profile at the edge of the contact zone, as it predicts a corner at right angle \textcolor{black}{while} the data exhibit a much smoother profile. This was already mentioned in a seminal work by Horn, Israelachvili and Fribac~\cite{Horn1987a}, and attributed to the flexural stiffness of the mica layer, that is not included in the model as solids are supposed homogeneous. \textcolor{black}{This non perfect flattening is therefore well known, but is highlighted in the present study because of the sub-pixel-detection of the fringe profile, and the smaller values of load and contact radius explored.} Nonetheless, we will keep the JKR fit as a first order determination of the contact radius, and we will see in the next section that this will provide a variation of the contact radius with the force that is consistently well described by the JKR model and an effective elastic modulus describing the layered system.

In order to obtain a reliable characterization of the geometry, all the images had to be rotated by the same angle (of the order of the degree) before this analysis, due to the fact that the camera is not perfectly aligned with the entrance slit of the spectrometer and so the raw image is not ideally symmetric. One image, corresponding to a situation when the surfaces are far from contact, is rotated by a given angle and the separation profile is fitted at all scales with a parabola. The values of rotation angles are scanned, and the optimum angle corresponds to the fit associated with the minimum sum of squared residuals.

Finally, $a$ is measured with a precision of~$0.03~\mathrm{\mu m}$ given by the standard deviation of the signal, and an accuracy of~$1~\mathrm{\mu m}$ due to the uncertainty on the value of~$R$ (mainly caused by the fact that the separation profile is observed up to a maximum scale~$z_\mathrm{max}\sim 50~\mathrm{nm}\ll R\sim 1~\mathrm{cm}$). This means that this method doesn't provide reliable values of $a$ when $a\lesssim 1~\mathrm{\mu m}$, which is typically the case for [C$_4$C$_1$Pyrr][NTf$_2$] under low loads.

In the literature, the measured force~$F$ is generally rescaled by the radius of curvature~$R$ to compute an equivalent surface energy~$F/R$, considering that mechanical deformations are negligible and that the Derjaguin approximation applies. In the opposite case when the surfaces are strongly flattened, it is reasonable to assume that the total force~$F$ is mainly due to the interaction in the flattened region, and so to rescale it by the contact area~$\pi a^2$ to compute the mean local pressure~$F/(\pi a^2)$. In this study, we explore a broad range of situations from non measurable deformation to strong deformations, that is why we have chosen to simply use the force~$F$ without any rescaling in the plots.

\section{Results and Discussion}
\label{sec:ResultsandDiscussion}

\subsection{Calibrations in dry atmosphere}
\label{subsec:Calibrationsindryatmosphere}

\begin{figure*}
\centering
\includegraphics{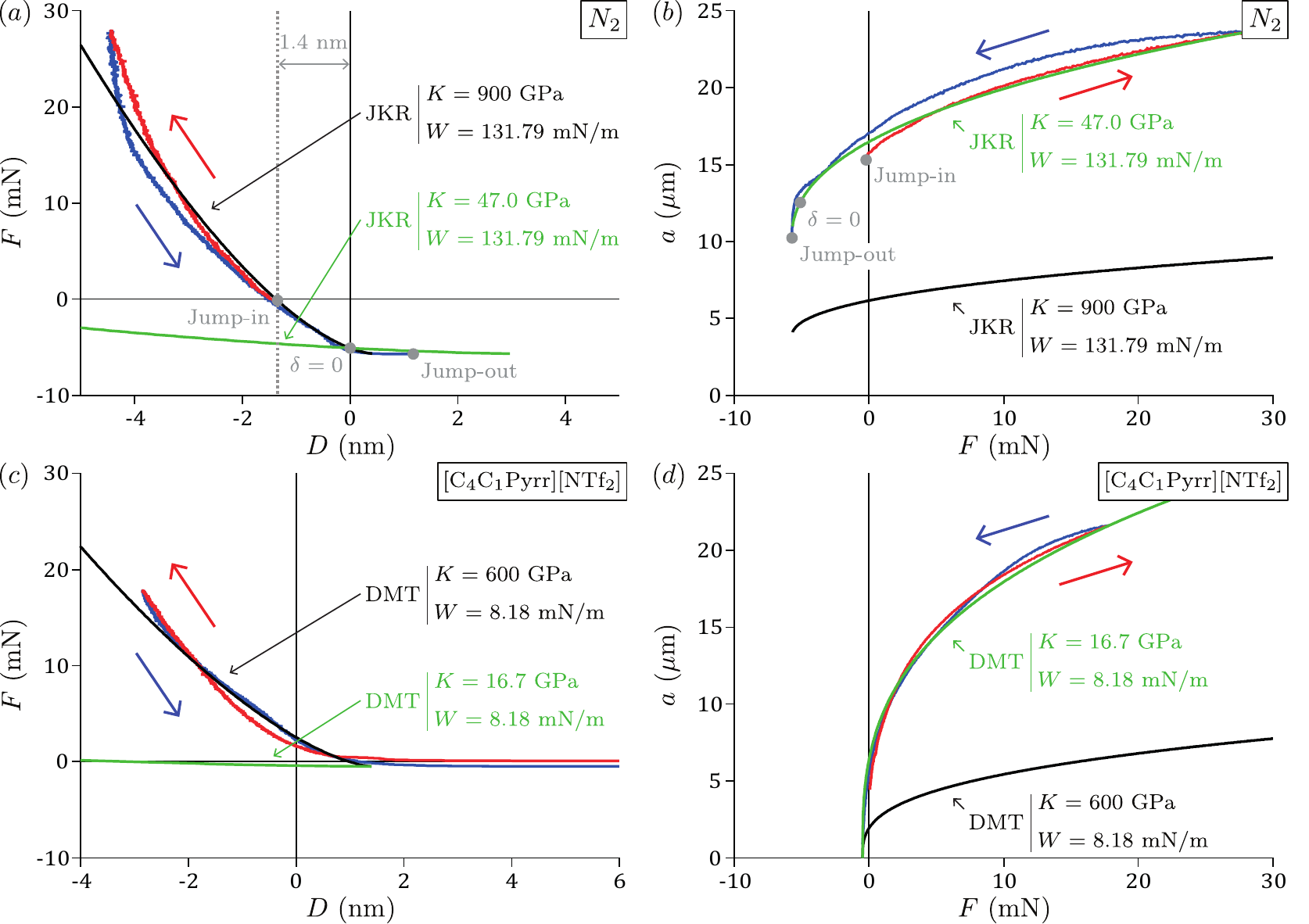}
\caption{(a)~Force~$F$ as a function of distance~$D$ and (b)~contact radius~$a$ as a function of force~$F$ when approaching (in red) and retracting (in blue) the top surface across N$_2$ with the stepper motor at~$v=13.2~\mathrm{nm/s}$. The green curve is a JKR fit of~$a(F)$ (equations~\ref{eq:JKR_F_a_delta_a}) with $K=47.0\pm 0.5~\mathrm{GPa}$, $W=131.79~\mathrm{mN/m}$ and $D_\mathrm{ref}=\delta+D=0$; the black curve is a JKR fit of~$F(\delta)$ (equations~\ref{eq:JKR_F_a_delta_a}) with $K=900\pm 200~\mathrm{GPa}$, $W=131.79~\mathrm{mN/m}$ and $D_\mathrm{ref}=\delta+D=0$. (c)~Force~$F$ as a function of distance~$D$ and (d)~contact radius~$a$ as a function of force~$F$ when approaching (in red) and retracting (in blue) the top surface across [C$_4$C$_1$Pyrr][NTf$_2$] with the stepper motor at~$v=10.5~\mathrm{nm/s}$. The green curve is a DMT fit of~$a(F)$ (equations~\ref{eq:DMT_F_a_delta_a}) with $K=16.7\pm 0.5~\mathrm{GPa}$, $W=8.18~\mathrm{mN/m}$ and $D_\mathrm{ref}=\delta+D=1.4~\mathrm{nm}$; the black curve is a DMT fit of~$F(\delta)$ (equations~\ref{eq:DMT_F_a_delta_a}) with $K=600\pm 200~\mathrm{GPa}$, $W=8.18~\mathrm{mN/m}$ and $D_\mathrm{ref}=\delta+D=1.4 \pm 0.4~\mathrm{nm}$. The curves measured on approach and retraction are not superimposed because of adhesion hysteresis (mainly due to viscoelasticity and plasticity of the glue) and mechanical imperfections of the set-up (poor linearity of the stepper motor, tiny rotations of the solids, etc.).}
\label{Fig3}
\end{figure*}

In Figures~\ref{Fig3}(a) and~\ref{Fig3}(b) are shown the force~$F$, the distance~$D$ and the contact radius~$a$ measured when approaching (in red) then retracting (in blue) the top surface with the stepper motor at $13.2~\mathrm{nm/s}$ across N$_2$. Initially separated by N$_2$, the surfaces are brought closer until they experience a strong van der Waals attraction, which together with the spring instability causes a jump-in to mica-mica contact and a slightly negative force ($F=-0.20~\mathrm{mN}$); thereafter the applied load is increased continuously and very large positive forces are reached. At some point ($F=27.91~\mathrm{mN}$), the direction of the motion is reversed, the applied load is decreased continuously and large negative forces are reached until the spring instability at $F_\mathrm{s}=-5.71~\mathrm{mN}$ leads to a jump-out to large distances.

The variation of the contact radius~$a$ with the force~$F$, plotted in Figure~\ref{Fig3}(b), is totally consistent with the work of Horn, Israelachvili and Fribac~\cite{Horn1987a}. Clearly, the jump-out happens at a non-zero value for the contact radius ($a_\mathrm{s}=10.05~\mathrm{\mu m}$), which is a typical feature of the JKR model. The measured $a(F)$ relationship is compared to the JKR prediction (equation~\ref{eq:JKR_F_a_delta_a}) in the following manner. First, the force $F_\mathrm{s}=-5.71~\mathrm{mN}$ reached just before the jump-out is used to compute the adhesion energy $W=-\frac{2 F_\mathrm{s}}{3\pi R}=131.79~\mathrm{mN/m}$. Then, a fitting procedure provides the elastic modulus $K=47.0\pm 0.5~\mathrm{GPa}$. As the range of a van der Waals attractive force across a gas is $d\sim 0.3~\mathrm{nm}$~\cite{Horn1987a}, a Maugis parameter $\mathrm{Ma}\sim19\gg 1$ is obtained from equation~\ref{eq:Maugis_Ma}, confirming the suitability of the JKR model for this system. The data are reasonably fitted by the model (see the curve in green), and the extracted $W$ and $K$ are consistent with the values reported in previous studies for similar systems~\cite{Horn1987a,McGuiggan2007a}. This validates \textit{a posteriori} the use of the JKR model to fit of the separation profile (as detailed in the previous section). Such agreement is fortunate, given JKR model supposes the solids are homogeneous, while the solids used in this experiment are complex stack of layers associated with different thicknesses, Young's moduli and Poisson's ratios. In this sense, the value of $K$ extracted from the fit is the elastic modulus of an homogeneous material that would give the same response than our complex system, i.e. it is an effective quantity. It thus makes sense that its value is between the Young's modulus of mica or glass and the Young's modulus of the glue (values from the literature given in Figure~\ref{Fig1}(c)). Of course, this effective quantity depends on the nature of the particular lenses used, and in particular on the mica and glue thicknesses that vary from one experiment to another. At last, we observe that the curves measured on approach and retraction are not superimposed, a phenomenon known as the adhesion hysteresis and due to non-elastic processes at play. For this system, it is generally assumed that adhesion hysteresis is mainly caused by viscoelasticity and plasticity in the glue layer~\cite{Horn1987a,McGuiggan2007a}.

The variation of the force~$F$ with the distance~$D$ is plotted in Figure~\ref{Fig3}(a). The curves measured on approach and retraction are not perfectly superimposed, mainly because of mechanical imperfections of the set-up, i.e the poor linearity of the stepper motor and tiny rotations of the solids during the loading-unloading procedure. We observe a clear change of~$D$ when the surfaces are in contact, of~$\sim5~\mathrm{nm}$ between the points of maximum load and of jump-out, or~$\sim2~\mathrm{nm}$ between jump-in and jump-out. \textcolor{black}{Potentially, imperfections of the double cantilever spring may induce a tiny rotation of the solids when varying the load, leading to a progressive dealignment of the light and so to a shift of the fringe positions and of the extracted distance. However, we ruled out this possible artefact by checking the tuning of the optics at regular intervals during the loading-unloading cycle. In addition, we have made sure that this phenomenon is not due to a potential contamination of a particular experiment, by systematically observing that mica undergoes a significant indentation in many separate experiments with different mica sheets.} As the mica is the single material separating the two silver mirrors, such variation can only be explained by a compression of the two mica layers in contact. In many previous studies with SFA, it was clearly stated~\cite{Israelachvili1978a,Horn1987a,Math2007a} or implicitly assumed that the mica is bending but experiences a negligible compression. This is mostly due to the fact that measurements of the fringe positions were performed by eye with thinner mica, respectively producing scattered points and reducing the amplitude of compression. In some studies with a so-called extended Surface Force Apparatus, compression of mica was mentioned but not studied specifically~\cite{EspinosaMarzal2012a,EspinosaMarzal2014a,Jurado2015a,Jurado2016a,Zachariah2016a,Zachariah2017a}. Such significant compression of the mica raises a technical difficulty for analyzing the FECO. Like what is done usually, we have supposed two mica layers of same constant thickness~$e_\mathrm{mica,0}$ separating a vacuum layer of thickness~$D$ (with known and constant refractive indexes), which is clearly wrong here when the surfaces are in contact. To be more rigorous, we have re-analyzed the FECO when the surfaces are in contact, supposing a single mica layer of variable thickness~$2e_\mathrm{mica}=2e_\mathrm{mica,0}-2\delta e_\mathrm{mica}$ ($\delta e_\mathrm{mica}$ defined positive for compression and negative for dilatation). In fact, the two methods provide the same amplitude of compression, because in both cases we are looking at a small variation of a thickness around a given reference point, and the FECO are not sensitive to the refractive index associated with this varying thickness, taken as the one of vacuum or the one of mica. Consequently, the variations of the distance~$D$ observed here have to be interpreted as changes of N$_2$ thickness when the surfaces are out of contact, and changes of mica thickness when the surfaces are in contact.

In these conditions, an important question is how can we define the mechanical origin, i.e the point at which $D=0$? A sensible choice is to select the point where the indentation of the surfaces is zero, i.e.~$\delta=0$. Following the JKR model, $\delta$ is positive at the maximum load (compression) and negative at the jump-out (dilatation). The point of zero indentation is located in between, and does not correspond to the point of zero force, i.e~$F=0$. Considering that the measured $a(F)$ relationship is reasonably fitted by the JKR model, we have taken the point~$\delta=0$ from the fit as the mechanical zero for the measured~$D$ (as indicated in Figures~\ref{Fig3}(a) and~\ref{Fig3}(b), corresponding to a mica thickness $e_\mathrm{mica,0}=7.431~\mathrm{\mu m}$) and leading to a distance~$D$ that can be negative. This procedure to find the mechanical origin is significantly different from what is done usually. In general, the force profile in dry atmosphere is not measured, but the surfaces are slowly approached until they jump-in to contact and it is at this point that the ``mica thickness'' is calibrated. In our experiment, the surfaces are already compressed by~$1.4~\mathrm{nm}$ just after the jump-in. The usual procedure therefore leads to an underestimation of the mica thickness by $\sim 1~\mathrm{nm}$ for this $\sim 7~\mathrm{\mu m}$-thick mica, and an equivalent outward shift of the force profile measured after injecting the liquid (adhesion is generally much smaller across a liquid than dry atmosphere). Qualitatively, the shift is expected to decrease if the mica is thinner or if the spring is stiffer, but the error will be present in any case. Taking this effect into account is particularly important when investigating aspects at the molecular scale. In the following, we present three examples of situations where this matters.
\begin{itemize}
\item Since the conception of the SFA, negative distances from $-0.2~\mathrm{nm}$ to $-1.3~\mathrm{nm}$ have been reported when two mica surfaces separated by water jump-in to contact~\cite{Tabor1969a,Israelachvili1978a,Homola1989a,Raviv2001a,Raviv2002a,Raviv2002b,Perkin2006a}. This has been attributed to the washing of gas molecules and organic contaminants (carbon compounds) that are spontaneously deposited on the mica surfaces in air~\cite{Poppa1971a}, and to the dissolution in water of the potassium ions initially present on the mica surfaces. As adhesion is typically 10 times smaller in water than in dry atmosphere, the mica is expected to be less compressed after the jump-in across water than during the calibration after the jump-in across dry atmosphere. Because compression of mica was not considered in these studies, the thickness of the contaminant layer is probably underestimated, and the dependence of this effect on the mica thickness and spring constant may explain -at least in part- the strong variability on the reported values.
\item In the case of molecular liquid giving rise to a structural force profile under confinement, a good accuracy on the distance~$D$ is needed in order to identify the absolute number of ordered layers composing the film (as illustrated in subsection titled ``Influence of surface deformations on structural force profile'').
\item The SFA can be used to determine the slip length associated to a flow of liquid in vicinity of a solid surface. By definition, the slip length is the distance between the hydrodynamic origin and the mechanical origin, that is why a sub-nanometric resolution on the mechanical zero is required to measure nanometric slip lengths~\cite{Chan1985a,Israelachvili1986a,Klein1993a,Campbell1996a,CottinBizonne2008a,Lhermerout2018b}. 
\end{itemize}

Finally, one can ask whether the JKR model can also fit the $F(D)$ relationship. If we use equation~\ref{eq:JKR_F_a_delta_a} and a reference~$D_\mathrm{ref}=\delta+D=0$ (the origin $D=0$ corresponds to $\delta=0$), with the values $W=131.79~\mathrm{mN/m}$ and $K=47.0~\mathrm{GPa}$ coming from the fit of the $a(F)$ relationship (green curve in Figure~\ref{Fig3}(b)), the model does not fit at all and predicts an indentation that varies much more than in the experiment (by~$\sim 40~\mathrm{nm}$ instead of~$\sim 5~\mathrm{nm}$ in the explored range of force, see green curve in Figure~\ref{Fig3}(a)). Qualitatively, this is because these effective parameters correspond to the indentation of the whole mica/glue/glass system, while here we measure the indentation of the mica only. \textcolor{black}{If we consider that the different layers composing the solids behave as springs in series, one expects that the ratio of the indentation of mica relative to the indentation of glue is of the order of the ratio of the Young’s modulus of mica relative to the Young’s modulus of glue, i.e. $3.4~\mathrm{GPa}/62~\mathrm{GPa}=5.5\%$, which is indeed of the same order of magnitude than the measured ratio of $5~\mathrm{nm}/40~\mathrm{nm}=12\%$.} If we now relax the parameter~$K$ to fit the $F(D)$ relationship (black curve in Figure~\ref{Fig3}(a)), it does not fit the $a(F)$ relationship (for the same reason given just before, see black curve in Figure~\ref{Fig3}(b)) and it provides a value $K=900 \pm 200~\mathrm{GPa}$ one order of magnitude larger than the elastic modulus of mica. As we are probing a contact zone of size $a\sim e_\mathrm{mica}$, the mica layers cannot be considered as semi-infinite solids (this would require $a\ll e_\mathrm{mica}$) and the JKR model does not apply. Qualitatively, the finite size of the mica layers cuts off the range of the elastic deformations, leading to an apparent stiffening of the solids compared to their bulk counterparts. Some analytical solutions exist for the opposite case of an infinitely thin elastic layer between two rigid solids~\cite{Reedy2006a,Borodich2019a} (applicable for $a\gg e_\mathrm{mica}$), but not for the intermediate case present here ($a\sim e_\mathrm{mica}$). \textcolor{black}{To summarize, the JKR model reasonably fits the variation of the contact radius with the force -suggesting that the whole mica/glue/glass system acts as an effective homogeneous material-, but completely fails to describe the variation of the force with the distance -because only the indentation of the mica is measured, which form a thin layer that cannot be considered as semi-infinite-.} In order to simultaneously fit the $a(F)$ and $F(D)$ relationships with a coherent set of parameters, complex models including the description of all the layers are needed, with solutions computed under some approximations~\cite{Math2007a} or by finite element methods~\cite{Sridhar1997a,McGuiggan2007a} .

\subsection{Contact mechanics across an ionic liquid}
\label{subsec:Contactmechanicswithanionicliquid}

In Figures~\ref{Fig3}(c) and~\ref{Fig3}(d) are shown the force~$F$, the distance~$D$ and the contact radius~$a$ measured when approaching (in red) then retracting (in blue) the top surface with the stepper motor at $10.5~\mathrm{nm/s}$ across [C$_4$C$_1$Pyrr][NTf$_2$]. When the ionic liquid is confined at the nanoscale, a structural force profile is observed, due to the organization of the ions in ordered layers. Initially far away, the surfaces are brought closer until they experience a repulsive wall at $D \sim 1.5~\mathrm{nm}$ (\textcolor{black}{thereafter reported as layer $i=2$}) for a load up to $F=0.39~\mathrm{mN}$, then a layer is squeezed-out and the surfaces jump-in to another repulsive wall at $\sim 0.5~\mathrm{nm}$ (\textcolor{black}{thereafter reported as layer $i=1$}); thereafter the applied load is increased continuously and very large positive forces are reached. At some point ($F=17.80~\mathrm{mN}$), the direction of the motion is reversed, the applied load is decreased continuously and small negative forces are reached until the spring instability at $F_\mathrm{s}=-0.47~\mathrm{mN}$ leads to a jump-out to large distances. In this subsection, we focus on the contact mechanics of the system when the liquid is composed of a single layer of ions (layer $i=1$). The influence of the mechanical deformations on the structural force profile will be detailed in the next subsection.

The variation of the contact radius~$a$ with the force~$F$ is plotted in Figure~\ref{Fig3}(d). Clearly, the jump-out happens at a zero value for the contact radius ($a_\mathrm{s}$ within the systematic experimental error), which is a typical feature of the DMT model. The measured $a(F)$ relationship is compared to the DMT prediction (equation~\ref{eq:DMT_F_a_delta_a}) in the following manner. First, the force $F_\mathrm{s}=-0.47~\mathrm{mN}$ reached just before the jump-out is used to compute the adhesion energy $W=-\frac{F_\mathrm{s}}{2\pi R}=8.18~\mathrm{mN/m}$. Then, a fitting procedure provides the elastic modulus $K=16.7\pm 0.5~\mathrm{GPa}$. The data are reasonably fitted by the model (see the curve in green), and \textcolor{black}{the extracted $K$ is of the same order than the value previously obtained from the analysis of the variation of the contact radius with the force in N$_2$}. The measurements across N$_2$ and [C$_4$C$_1$Pyrr][NTf$_2$] have been performed with the same lenses but it cannot be ensure that the spots used on the surfaces are strictly the same, that is why a different glue thickness could explain why the effective elastic modulus changed by a factor 3 after injecting the liquid. Our findings disagree with Christenson who stated that SFA measurements are always performed in the JKR regime, based on the calculation of the Maugis parameter~$\mathrm{Ma}$ from equation~\ref{eq:Maugis_Ma}~\cite{Christenson1996a}. The difficulty of such approach is that $\mathrm{Ma}$ strongly depends on the range~$d$ of the attractive forces, which can be delicate to estimate. For this measurement, we have to assume $d\gtrsim 5~\mathrm{nm}$ for the range of the attractive force across the ionic liquid, to get from equation~\ref{eq:Maugis_Ma} a Maugis parameter $\mathrm{Ma}\lesssim0.4$ that corresponds to the DMT regime. Looking at the value of the contact radius just before the jump-out instead is an extremely sensitive method, that does not rely on any microscopic parameters. For our experiment with the ionic liquid, JKR model predict that the surfaces would separate at $a_\mathrm{s}=\left(\frac{3\pi R^2 W}{2K}\right)^{1/3}=6.39~\mathrm{\mu m}$ (for $W=-\frac{2F_\mathrm{s}}{3\pi R}=10.91~\mathrm{mN/m}$ and $K=16.7~\mathrm{GPa}$) while we clearly observe that the jump-out happens at a contact radius that is below the systematic experimental error of $1~\mathrm{\mu m}$, unambiguously showing that we are in the present case not in the JKR regime but in the DMT regime.

Two situations have been addressed in the seminal paper of Horn, Israelachvili and Fribac~\cite{Horn1987a}: the case of strong adhesion ($W\gtrsim 100~\mathrm{mN/m}$) over a range of a fraction of nanometer that was obtained with a dry atmosphere and was well fitted by the JKR model, and the case of negligible adhesion ($W\lesssim 1~\mathrm{mN/m}$) that was obtained with an aqueous electrolyte and was well fitted by the Hertz model. For our intermediate situation of moderate adhesion ($W\sim 10~\mathrm{mN/m}$) over a range of a few nanometers obtained with an ionic liquid, \textcolor{black}{we have shown that the variation of the contact radius with the force is well fitted by the DMT model}. Such situation of moderate adhesion over a range of a few nanometers is not specific to ionic liquids only, but is frequently encountered in SFA experiments, for example with apolar liquids, salt solutions, polymer melts or liquid crystals. Our findings are thus of general interest and have potentially important consequences, because using the correct model of contact mechanics is crucial to interpret force measurements, in particular for the two situations listed below.
\begin{itemize}
\item The jump-out force $F_\mathrm{s}$ obtained with force measurement techniques is routinely used to deduce the surface energy~$W$. As the relationship between these two quantities depends on the model ($W=-\frac{2F_\mathrm{s}}{3\pi R}$ in JKR model, $W=-\frac{F_\mathrm{s}}{2\pi R}$ in DMT model), it is crucial to know the regime of contact in order to extract reliable values~\cite{Grierson2005a}.
\item When investigating friction in the boundary lubrication regime with smooth adhering surfaces, the applied load is in general large enough to flatten the sliding surfaces. These mechanical deformations have to be known in order to interpret the data, in particular to determine whether the friction is controlled by the area of contact or by the load, and to unravel the role of adhesion~\cite{Derjaguin1988a,Homola1989a,Israelachvili1994a,Kumacheva1998a,Berman1998a,Bogdanovic2001a,Grierson2005a,Bureau2010a,Lessel2013a}. 
\end{itemize}

The variation of the force~$F$ with the distance~$D$ is plotted in Figure~\ref{Fig3}(c). When the surfaces are separated by a single layer of ions (layer $i=1$), we clearly observe that~$D$ can be negative and changes by~$\sim~4\mathrm{nm}$ between the points of maximum load and of jump-out. This is due to the compressibilities of the mica layers and of the liquid film, i.e the materials separating the two silver mirrors. As explained in the previous subsection, for such small changes of~$D$ it is not possible to disentangle the contributions coming from the mica (indentation~$2e_\mathrm{mica,0}-2e_\mathrm{mica}=2\delta e_\mathrm{mica}$, defined positive for compression and negative for dilatation) and the liquid (thickness~$D_\mathrm{liquid}$) with the FECO. In fact, we effectively measure the distance between the silver mirrors~$D_\mathrm{mirrors}=D_\mathrm{liquid}+2e_\mathrm{mica}$, from which we subtract the undeformed mica thickness $2e_\mathrm{mica,0}$ calibrated in dry atmosphere, to finally obtain:

\begin{equation}
D=D_\mathrm{mirrors}-2e_\mathrm{mica,0}=D_\mathrm{liquid}-2\delta e_\mathrm{mica}~\textrm{.}
\label{eq:definition_D}
\end{equation}

\noindent One can ask whether the DMT model can also fit the $F(D)$ relationship, supposing that most of the change of~$D$ comes from the indentation of the mica. If we use equation~\ref{eq:DMT_F_a_delta_a} with the values $W=8.18~\mathrm{mN/m}$ and $K=16.7~\mathrm{GPa}$ coming from the fit of the $a(F)$ relationship (green curve in Figure~\ref{Fig3}(d)), the model does not fit at all and predicts an indentation that varies much more than in the experiment (by~$\sim 50~\mathrm{nm}$ instead of~$\sim 4~\mathrm{nm}$ in the explored range of force, see green curve in Figure~\ref{Fig3}(c)). If we now relax the parameter~$K$ to fit the $F(D)$ relationship (black curve in Figure~\ref{Fig3}(c)), it does not fit the $a(F)$ relationship (black curve in Figure~\ref{Fig3}(d)) and it provides a value $K=600 \pm 100~\mathrm{GPa}$ one order of magnitude larger than the elastic modulus of mica. Similarly to the calibrations in dry atmosphere, this is because we do not measure the indentation due to the layers behind the silver mirrors (glue and glass), and a suitable model of contact mechanics should include the finite thickness of the mica. Under the assumption that most of the change of~$D$ comes from the indentation of the mica only, the role of the liquid layer is included implicitly in the adhesion energy, and explicitly in the reference~$D_\mathrm{ref}=\delta+D=1.4 \pm 0.4~\mathrm{nm}$ that was simultaneously fitted (black curve in Figure~\ref{Fig3}(c)) and also used for the first comparison (green curve in Figure~\ref{Fig3}(c)).

\subsection{Influence of surface deformations on structural force profile}
\label{subsec:Influenceofsurfacedeformationsonstructuralforceprofile}

\begin{figure}
\centering
\includegraphics{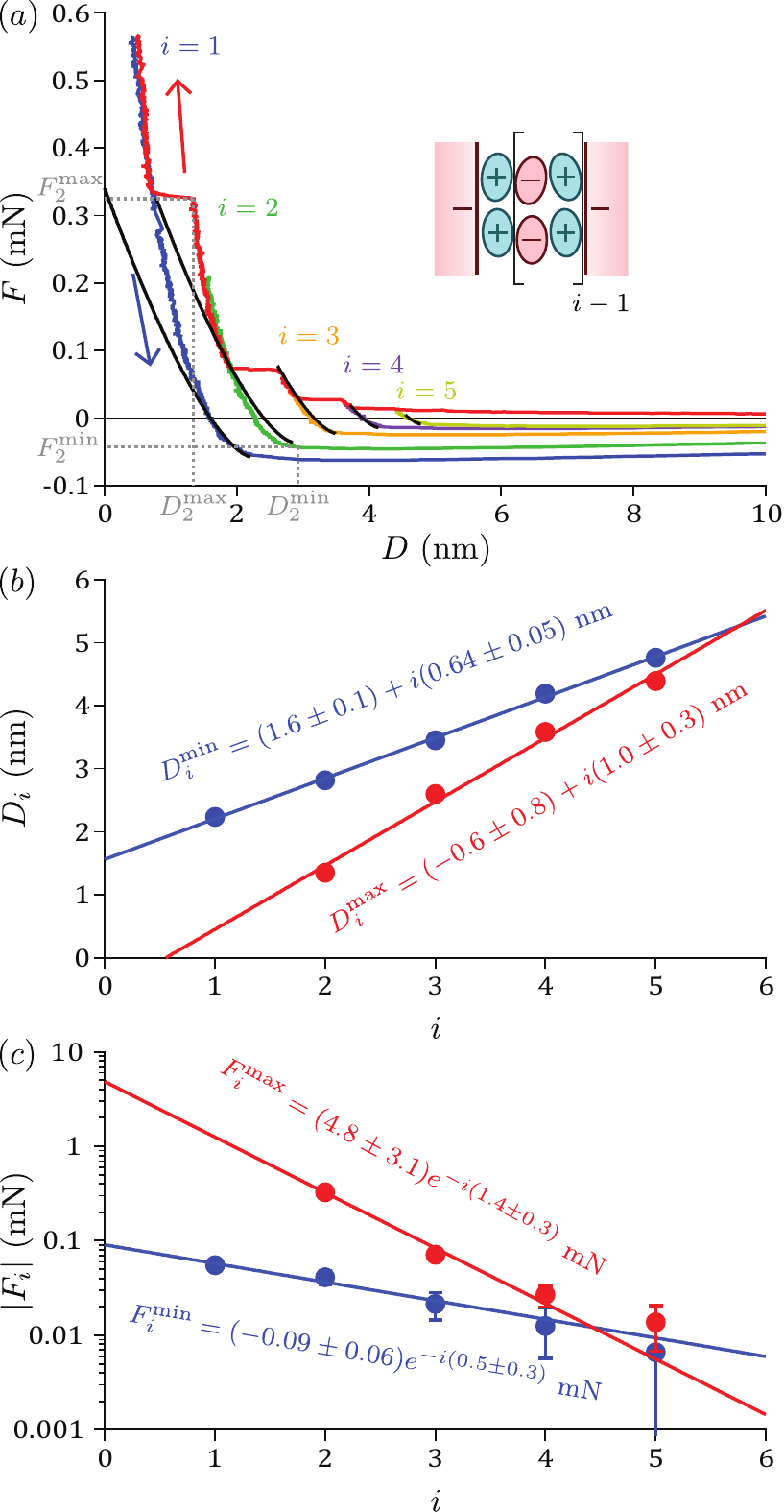}
\caption{(a)~Force profile measured with [C$_4$C$_1$Pyrr][NTf$_2$] between mica surfaces when approaching or retracting the top surface with the piezoelectric tube at~$v=0.5~\mathrm{nm/s}$, showing structuring with 5 distinguishable layers labeled by $i$. For clarity, only the full approach is shown (in red), together with retractions from layers~$i=1$ (in blue), $i=2$ (in green), $i=3$ (in orange), $i=4$ (in purple), $i=5$ (in yellow). The black lines are the fit with equation~\ref{eq:structural_force_DMT_fit}. Inset: proposed structure of alternating layers of cations and anions. (b)~Distances~$D_i$ measured at the points of maximum and minimum force (respectively in red and blue), as a function of the layer index~$i$. Straight lines are the corresponding linear fits (equations~\ref{eq:D_i_fits}). (c)~Forces~$\left|F_i\right|$ measured at the points of maximum and minimum force (respectively in red and blue), as a function of the layer index~$i$, in log-lin representation. Straight lines are the corresponding exponential fits (equations~\ref{eq:F_i_fits}).}
\label{Fig4}
\end{figure}

In this subsection, we now focus on the detailed shape of the structural force profile. Because of the spring instability, the surfaces experience a jump-in on approach every time a layer is squeezed-out, an a jump-out on retraction from a given layer. Therefore several runs are necessary for the most comprehensive exploration of the force profile. Supplementary Figure~1 shows the force profiles measured when approaching the top surface up to a given layer and retracting from this layer with the piezoelectric tube at~$v=0.5~\mathrm{nm/s}$. Five layers can be distinguished, and are labeled by $i$. From run to run, the whole force profile randomly shifts laterally by a fraction of nanometer. \textcolor{black}{As the jump-in distances and the forces are reproducible, we consider that these shifts are non physical, but result from imperfections of the set-up like tiny rotations of the solids or fluctuations of the contact spots on the surfaces that lead to slight dealignement of the light}. That is why we have shifted manually the force profiles such that all the approaches are superimposed on the approach up to layer~$i=1$ (in red). In Figure~\ref{Fig4}(a) is shown the resulting force profile with for clarity only the approach up to layer~$i=1$ (in red), and the retractions from the different layers ($i=1$ in blue, $i=2$ in green, $i=3$ in orange, $i=4$ in purple, $i=5$ in yellow). Such structural force profile has been observed many times with ionic liquids, and attributed to the ordering of ions in the film, with a structure consisting of alternating layers of anions and cations (as sketched in inset)~\cite{Horn1988a,Atkin2007a,BouMalham2010a,Perkin2010a,Ueno2010a,Zhang2012a,Hoth2014a,Cheng2015a,Garcia2017a}. For a given layer~$i$, the distance~$D$ is not constant. In general, this is interpreted as a result of the change of the local liquid density for infinitely stiff surfaces, and the structural force profile is fitted with a semi-empirical exponentially decaying harmonic function of the form:

\begin{equation}
F(D)=F_0\exp\left(-\frac{D-D_0}{\zeta}\right)\cos\left(2\pi\frac{D-D_0}{\lambda}\right) \textrm{~,}
\label{eq:structural_force_exp_cos_fit}
\end{equation}

\noindent where the 4 fitting parameters are the period of the oscillations $\lambda$, the decay length $\zeta$, and the position $D_0$ and amplitude $F_0$ of the first layer. In the following, we show that our data does not follow this description, because of the mica compression that is dominant compared to the liquid compression. To characterize the structural force profile, we have measured for each layer~$i$ the distances~$D_i^\mathrm{max}$, $D_i^\mathrm{min}$ and the forces~$F_i^\mathrm{max}$, $F_i^\mathrm{min}$ at the points of maximum and minimum force (i.e. respectively just before the jump-in and jump-out, as indicated in Figure~\ref{Fig4}(a) for $i=2$).

Figure~\ref{Fig4}(b) shows the variations of the distances~$D_i^\mathrm{max}$, $D_i^\mathrm{min}$ with the layer index~$i$. The two curves exhibit a good linearity, and are fitted with the relations:

\begin{equation}
\left\{
\begin{array}{lll}
D_i^\mathrm{max}&=&D_0^\mathrm{max}+i\lambda^\mathrm{max} \\[5pt]
D_i^\mathrm{min}&=&D_0^\mathrm{min}+i\lambda^\mathrm{min}
\end{array}
\right.\textrm{~,}
\label{eq:D_i_fits}
\end{equation}

\noindent where the slopes $\lambda^\mathrm{max}$, $\lambda^\mathrm{min}$ represent the mean layer thickness and the intersects $D_0^\mathrm{max}$, $D_0^\mathrm{min}$ correspond to the position of the extrapolated layer $i=0$ (fitted values indicated in the Figure). The mean layer thickness obtained from the maxima $\lambda^\mathrm{max}=1.0\pm 0.3~\mathrm{nm}$ is significantly larger than the mean layer thickness obtained from the minima $\lambda^\mathrm{min}=0.64\pm 0.05~\mathrm{nm}$. This surprising observation, that has been noticed in the past~\cite{Lhermerout2018b}, can in fact be rationalized by considering the influence of the mica compression. According to the DMT model for a given layer, the indentation of the solids is zero at the minimum force (jump-out point) and continuously increases up to the maximum force (jump-in point). When going from layer~$i=1$ to~$i=5$, the range of force explored decreases (because of repulsive maxima and adhesive minima that come closer to $F=0$), as well as the amplitude of mica compression. The variation of~$D_i^\mathrm{max}$ with~$i$ thus includes a systematic decrease of the mica compression, leading to an overestimation of the mean layer thickness. On the contrary, the variation of~$D_i^\mathrm{min}$ with~$i$ does not include any influence of the mica compression, and provides the true mean layer thickness $0.64 \pm 0.05~\mathrm{nm}$. Another method to determine the mean layer thickness consists in measuring the average jump-in distance, supposing an unchanged mica compression and a fast viscous relaxation during the squeeze-out events. This provides a consistent value of $0.64 \pm 0.01~\mathrm{nm}$, confirming our interpretation of the effect of the mica compression on the mean layer thickness. Interestingly, this value is smaller than the mean ion pair diameter of $0.79~\mathrm{nm}$ (given by $\left(\frac{M}{\rho N_\mathrm{A}}\right)^{1/3}$ with $M$ the molar mass of the ionic liquid, $\rho$ its bulk density and $N_\mathrm{A}$ the Avogadro's number~\cite{Horn1988a}), perhaps suggesting a denser packing of ions in confinement than in the bulk. However, our value is also smaller than the previous measurements performed with the same ionic liquid, reporting a mean layer thickness of $0.80 \pm 0.04~\mathrm{nm}$ between two mica surfaces with a SFB~\cite{Smith2017a} and $0.79~\mathrm{nm}$ between a mica surface and a Si$_3$N$_4$ tip with an AFM~\cite{Hayes2009a}. A possible explanation for this difference is the inherent contribution from viscosity to the force profile, in particular in vicinity to the jump-in and jump-out instability. For the method using the positions of the minima, viscosity tends to move the point of minimum force towards larger distances even more than adhesion is larger; for the method using the jump-in distances, viscosity tends to reduce the jump-in distances. In both cases, viscous effects possibly lead to an underestimation of the mean layer thickness. Previous studies may be less affected by viscosity, as retractions were performed by slow steps in the SFB study~\cite{Smith2017a} (with similar radius of curvature), and the radius of curvature was six orders of magnitude smaller in the AFM study~\cite{Hayes2009a} (with a velocity less than an order of magnitude larger). Therefore, applying a continuous motion to the surfaces is necessary to measure the detailed shape of the force profile, but applying a stepped motion is preferable to accurately determine the mean layer thickness.

Figure~\ref{Fig4}(c) shows the variations of the forces~$\left|F_i^\mathrm{max}\right|$, $\left|F_i^\mathrm{min}\right|$ with the layer index~$i$. In this log-lin representation, the two curves exhibit a good linearity, and are exponentially fitted with the relations:

\begin{equation}
\left\{
\begin{array}{lll}
F_i^\mathrm{max}&=&F_0^\mathrm{max}\exp\left(-i\frac{\lambda^\mathrm{max}}{\zeta^\mathrm{max}}\right) \\[5pt]
F_i^\mathrm{min}&=&F_0^\mathrm{min}\exp\left(-i\frac{\lambda^\mathrm{min}}{\zeta^\mathrm{min}}\right)
\end{array}
\right.\textrm{~,}
\label{eq:F_i_fits}
\end{equation}

\noindent where the slopes give access to the ratios $\frac{\lambda^\mathrm{max}}{\zeta^\mathrm{max}}$, $\frac{\lambda^\mathrm{min}}{\zeta^\mathrm{min}}$ of the period of the oscillation on the decay length, and the intersects correspond to the amplitudes $F_0^\mathrm{max}$, $F_0^\mathrm{min}$ of the extrapolated layer $i=0$ (fitted values indicated in the Figure). The curves obtained with the maxima and the minima are clearly distinct, as a consequence of the asymmetry of the envelope of the force profile with the horizontal axis. Such asymmetry cannot be due to the van der Waals contribution which is always attractive for symmetric systems. It cannot be explained by the anomalously long-range electrostatic force that has been observed with concentrated electrolytes~\cite{Gebbie2015a,Smith2016a}, because exponentially fitting the mean amplitude $\left(F_i^\mathrm{max}+F_i^\mathrm{min}\right)/2$ gives a decay of $\sim 1~\mathrm{nm}$ and an amplitude of $\sim 4~\mathrm{mN}$, respectively one order of magnitude smaller and two orders of magnitude larger than for long-range electrostatic force reported for this system~\cite{Smith2016a,Lhermerout2018b}. In fact, the exponentially decaying harmonic oscillation given by equation~\ref{eq:structural_force_exp_cos_fit} can be predicted theoretically in the asymptotic limit of large distances (far-field term), and an additional (non-oscillating) exponentially decay has been proposed as a correction at small distances (short-field term, with 2 additional fitting parameters)~\cite{Hoth2014a,Moazzami-Gudarzi2016a,Schon2018a}. This second term, that is intrinsic to the liquid for infinitely stiff surfaces, could contribute to the asymmetry of the measured force profile. The significant deformations of the surfaces could also contribute to the asymmetry, in the following manner. According to the DMT model for a given layer, the surfaces are not deformed at the minimum force (jump-out point), and flattened at the maximum force (jump-in point). Even if the force profile was symmetric for equivalent undeformable surfaces (no short-field term), the flattening at the maximum force would intuitively tend to make the squeeze-out harder, leading to a maximum that is larger than the minimum in absolute value. When going from layer~$i=1$ to~$i=5$, the range of force explored decreases, as well as the amplitude of flattening, and the points of maximum and minimum forces are more and more symmetric around the horizontal axis.

An important aspect to interpret structural force profiles is to identify the composition of the layers. As the period is similar to the mean ion pair diameter, it is qualitatively considered that one squeeze-out event corresponds to the squeeze-out of an electroneutral ``slab'' of one cation layer and one anion layer. In the case of negatively charged surfaces, the first layer ($i=1$) is then assumed to be composed of a monolayer of cations (as sketched in inset of Figure~\ref{Fig4}(a)). Direct solid-solid contact is never reached, because of the strong electrostatic attraction between the cations and the negatively charged surfaces. For our experiment, one can ask whether we really reach this single layer of cations within the explored range of loads. The position of the first layer at the point of minimum force (including no indentation of the surfaces) is $D^\mathrm{min}_1=2.2\pm1.0~\mathrm{nm}$, a bit larger than the cation sizes (given in Figure~\ref{Fig1}(a)). However, the measurement of the absolute distance~$D$ depends on many delicate steps (alignment of the optics, calibration of the mica thickness in dry atmosphere, choice of a particular run to shift the force profiles laterally), and deducing the thickness of the monolayer from the cation sizes requires to know their conformations, making accurate comparisons difficult. If we extrapolate the exponential fit of $\left|F_i^\mathrm{max}\right|$ to the layer $i=1$ (see Figure~\ref{Fig4}(c)), it predicts that the next squeeze-out event would take place at a force $F=1.3 \pm 0.6~\mathrm{mN}$, while we do not observe any additional jump-in for a force up to $F=17.80~\mathrm{mN}$ (see Figure~\ref{Fig3}(c)). Thus, we think that the layer seen at the maximum load is indeed composed of a single layer of cations, and we identify it as $i=1$.

In the SFA literature, the structural force profile is generally fitted with an exponentially decaying harmonic oscillation (equation~\ref{eq:structural_force_exp_cos_fit}, with 4 fitting parameters), neglecting surface deformations, which seems reasonable because the eye measurements produce scattered points and/or the retraction branches are not explored. However, it has been pointed out that the oscillation is not rigorously sinusoidal, already in the first paper reporting a structural force for a simple liquid~\cite{Horn1981a}, and even more clearly in a recent study with an extended Surface Force Apparatus~\cite{Zachariah2016a}. In Supplementary Figure~2, we have plotted two exponentially decaying harmonic oscillation (equation~\ref{eq:structural_force_exp_cos_fit}), the one in gray corresponding to the parameters $F_0=F_0^\mathrm{max}$, $D_0=D_0^\mathrm{max}$, $\zeta=\zeta^\mathrm{max}$ and $\lambda=\lambda^\mathrm{max}$, and the one in black corresponding to the parameters $F_0=F_0^\mathrm{min}$, $D_0=D_0^\mathrm{min}$, $\zeta=\zeta^\mathrm{min}$ and $\lambda=\lambda^\mathrm{min}$. None of these curves fit the measured force profile, and adding an exponential, short-field, term would not improve the situation. For an exponentially decaying harmonic oscillation, the stable branches are half convex and half concave, each minimum is located almost at the middle position between the surrounding maxima, and one value of~$D$ corresponds to a single value of~$F$ (force profile given by a function). In the experiment, the stable branches are always convex, the minima and maxima are not regularly spaced, and one value of~$D$ corresponds to a several value of~$F$ (we can have $D_i^\mathrm{min}>D_{i+1}^\mathrm{max}$). So it is clear that an exponentially decaying harmonic function is not appropriate to describe the detailed shape of the measured structural force profile. We propose a heuristic description, assuming on the contrary that the mica compression dominates the liquid compression, i.e. that the change of~$D$ within each layer comes only from the indentation of mica (elastic modulus~$K$). For each layer, we suppose that the mica indentation is given by the DMT model (equation~\ref{eq:DMT_F_a_delta_a}) with a reference $D_{i,\mathrm{ref}}=\delta_i+D=D_0^\mathrm{min}+i\lambda^\mathrm{min}$ and an adhesion $W_i=\left(-\frac{F_0^\mathrm{min}}{2\pi R}\right)\exp\left(-i\frac{\lambda^\mathrm{min}}{\zeta^\mathrm{min}}\right)=W_0\exp\left(-i\frac{\lambda^\mathrm{min}}{\zeta^\mathrm{min}}\right)$, up to a maximum force $F_0^\mathrm{max}\exp\left(-i\frac{\lambda^\mathrm{max}}{\zeta^\mathrm{max}}\right)$. Finally, we obtain the following expression:

\begin{equation}
\begin{array}{l}
F_i(D)=KR^{1/2}\left(D_0^\mathrm{min}+i\lambda^\mathrm{min}-D\right)^{3/2}\\[10pt]
\qquad\qquad -2\pi R W_0\exp\left(-i\frac{\lambda^\mathrm{min}}{\zeta^\mathrm{min}}\right)\\[10pt]
\textrm{for}~D\leq D_0^\mathrm{min}+i\lambda^\mathrm{min}~\textrm{and}~F_i\leq F_0^\mathrm{max}\exp\left(-i\frac{\lambda^\mathrm{max}}{\zeta^\mathrm{max}}\right)
\end{array}
\textrm{~.}
\label{eq:structural_force_DMT_fit}
\end{equation}

\noindent Basically, the liquid controls the positions (distance at the point of minimum force) and the strengths (forces at the points of minimum and maximum distances) of the layers, while the mica controls the shape of the profile within the layers. Note that the adhesion energies deduced from the DMT model coincide with ones that would be obtained supposing that the Derjaguin approximation applies, i.e. $W_i=-\frac{F_i^\mathrm{min}}{2\pi R}$, because the surfaces are not deformed at the point of minimum force in the DMT model, and supposed undeformable in the context of the Derjaguin approximation. Excluding the radius of curvature~$R$ that is measured independently, the force profile is described by 7 fitting parameters, which is the exact number of parameters required to describe an asymmetric structural force profile with deformable surfaces characterized by a single elastic modulus. To compare with our measurements, we have used the values $D_0^\mathrm{min}$, $\lambda^\mathrm{min}$ coming from the linear fit of $D_i^\mathrm{min}$, the values $F_0^\mathrm{min}$, $\zeta^\mathrm{min}$ coming from the exponential fit of $\left|F_i^\mathrm{min}\right|$, the values $F_0^\mathrm{max}$, $\zeta^\mathrm{max}$ coming from the exponential fit of $\left|F_i^\mathrm{max}\right|$, and we have fitted the remaining parameter~$K$. The fit shown in Figure~\ref{Fig4}(a) is very good at low loads, with $K=40\pm 5~\mathrm{GPa}$ remarkably close to the elastic modulus of solids that would be only composed of mica (value given in Figure~\ref{Fig1}(c)). In fact, when the force is close enough to the adhesion minimum, i.e. $\left|F_i-2\pi R W_i\right|\lesssim 0.1~\mathrm{mN}$, a contact zone of size $a\lesssim 3~\mathrm{\mu m} \ll e_\mathrm{mica}$ is probed, that is why the elastic deformations are affecting only the top mica layers, which can be considered as semi-infinite in these conditions~\cite{Sridhar1997a,McGuiggan2007a}. At larger loads, the system enters in a regime where the finite size of the mica layers cuts off the range of the elastic deformations, leading to an apparent stiffening of the solids, i.e. a larger gradient of the force profile and a unphysically large elastic modulus when abusively fitting with a model that supposes semi-infinite solids (see the two previous subsections). Contrary to Christenson~\cite{Christenson1996a}, we find for our system that the surface deformations have a strong influence on the force profile even at low loads. In particular, we expect this effect to be more important and independent of the mica thickness at low loads; and smaller and reduced for thinner mica at large loads.

For our specific system, we have shown that the detailed shape of the structural force profile is strongly affected by the mechanical deformations of the surfaces, with a mica compression that is dominant compared to the ionic liquid compression. For any system in general, the exponentially decaying harmonic oscillation due to local variations of liquid density is expected to be convoluted with the mechanical response of the confining solids. \textcolor{black}{This effect can \textit{a priori} be present not only with SFA/SFB, but also with AFM. In a typical AFM experiment (see for example~\cite{Hoth2014a}), the deflection~$\Delta$ of a cantilever is measured as a function of the approach position~$z$ imposed to the base of this cantilever. The force is deduced by multiplying the deflection by the calibrated spring constant, and the distance between the tip and the substrate is calculated as the difference between the cantilever deflection and the linear fit of the~$\Delta(z)$ relationship when the surfaces are in contact. In fact, this standard procedure supposes ideally that the surfaces are infinitely stiff in the fitting region. In practice real solids are compliant, and this method leads to subtracting the indentation of the surfaces, linearized in the fitting region. As recalled in subsection~\ref{subsec:Modelsofcontactmechanics}, models of contact mechanics generally predict that the relationship between the force and the indentation is not linear, and the calculated distance is not exactly equal to the distance between the tip and the substrate. In addition, the amplitude of the force depends on the geometry of the surfaces, and is therefore affected by their deformation. For these two reasons, the mechanical deformations of the confining surfaces are expected to influence the measured force profile also in the case of AFM.} It is very important to know the degree of convolution, i.e. whether the solids compression is negligible or dominant compared to the liquid compression, in order to interpret properly the structural force profile regarding the compressibility of the layers~\cite{Horn1981a,Smith2013b,Jurado2015a,Griffin2017a,Garcia2017a,Cheng2018a,Lhermerout2019a}. \textcolor{black}{Indeed, our study shows that the finite slope of the structural force profile in each layer is not necessarily due to a change with load of packing fraction or structure of the molecules in the structured film. In particular, this question of the ``elasticity'' of a thin liquid film is connected to a strong debate in the community, to understand how a liquid can exhibit a solid-like behaviour in nanoconfinement~\cite{Klein1995a,Khan2010a,Seddon2014a,Khan2016a,Comtet2017a}.} We propose a general criterion to distinguish the two opposite regimes of convolution, for a generic system exhibiting a structural force profile. We consider the mechanical response of a liquid confined between infinitely stiff solid surfaces (taken as an exponentially decaying harmonic oscillation $F_\mathrm{osci}(D)$), the mechanical response of two deformable solid surfaces in direct contact (assumed to be a simple Hertz force $F_K(D)$), the two responses being measured with an external spring (imposing a restoring spring force $F_k(D)$):

\begin{equation}
\left\{
\begin{array}{lll}
F_\mathrm{osci} (D)&=&-2\pi R W_0\exp\left(-\frac{D-D_0}{\zeta}\right)\cos\left(2\pi\frac{D-D_0}{\lambda}\right) \\[5pt]
F_K (D)&=&KR^{1/2}D^{3/2} \\[5pt]
F_k (D)&=&k\left[D-\left(D(t=0)-vt\right)\right]
\end{array}
\right.\textrm{~.}
\label{eq:forces}
\end{equation}

\noindent The pure liquid response is measured without spring instability if the gradient of the oscillating force is smaller than the gradient of the restoring spring force: $\left|\frac{\mathrm{d}F_\mathrm{osci}}{\mathrm{d}D}\right| < \left|\frac{\mathrm{d}F_\mathrm{k}}{\mathrm{d}D}\right|$. Similarly, the convoluted response of the liquid and the solids is only weakly affected by the solids if the gradient of the oscillating force is smaller than the gradient of the Hertz force: $\left|\frac{\mathrm{d}F_\mathrm{osci}}{\mathrm{d}D}\right| < \left|\frac{\mathrm{d}F_\mathrm{K}}{\mathrm{d}D}\right|$. For convenience, we define two dimensionless parameters~$N_k$ and~$N_K$ as the ratios of these gradients, and we estimate them with simple scalings:

\begin{equation}
\left\{
\begin{array}{lll}
N_k&=&\left|\frac{\mathrm{d}F_\mathrm{osci}}{\mathrm{d}D}\right| \bigg/ \left|\frac{\mathrm{d}F_\mathrm{k}}{\mathrm{d}D}\right| \sim 4\pi^2\frac{R}{k}\frac{W_0}{\lambda} \\[5pt]
N_K&=&\left|\frac{\mathrm{d}F_\mathrm{osci}}{\mathrm{d}D}\right| \bigg/ \left|\frac{\mathrm{d}F_\mathrm{K}}{\mathrm{d}D}\right| \sim \frac{8\pi^2}{3}\frac{R^{1/2}}{K}\frac{W_0}{\lambda^{3/2}}
\end{array}
\right.\textrm{~.}
\label{eq:N_k_N_K}
\end{equation}

\noindent The different parameters have analogous roles, even if the exact exponents and numerical factors are not the same. There is no spring instability (resp. small influence of surfaces deformations) when $N_k<1$ (resp. $N_K<1$), which is fulfilled for ``soft'' systems with small adhesion~$W_0$ and large period~$\lambda$, measured with a spring of large stiffness~$k$ (resp. with solids of large elastic modulus~$K$) and -less intuitively- with surfaces of small radius of curvature~$R$. In the following, we test these criteria with studies using different systems and instruments.

\begin{itemize}
\item For this SFB study with an ionic liquid ($R\sim 1~\mathrm{cm}$, $k\sim 3000~\mathrm{N/m}$, $K\sim 50~\mathrm{GPa}$, $W_0\sim 1~\mathrm{mN/m}$, $\lambda\sim 0.6~\mathrm{nm}$), we get $N_k\sim 2\cdot10^{2}$ and $N_K\sim 4$, in agreement with the fact that we have spring instabilities and a strong effect of the surface deformations on the structural force profile.
\item For previous AFM studies with ionic liquids~\cite{Atkin2007a,Hayes2009a,Hoth2014a} ($R\sim 20~\mathrm{nm}$, $k\sim 0.1~\mathrm{N/m}$, $K\sim 50~\mathrm{GPa}$, $W_0\sim 50~\mathrm{mN/m}$, $\lambda\sim 0.8~\mathrm{nm}$), we get $N_k\sim 5\cdot10^{2}$ and $N_K\sim 2\cdot10^{-1}$, in agreement with the fact that they have spring instabilities but a little effect of the surface deformations on the structural force profile.
\item For a previous SFA study with liquid crystals~\cite{Cross2004a} ($R\sim 0.3~\mathrm{cm}$, $k\sim 2000~\mathrm{N/m}$, $K\sim 50~\mathrm{GPa}$, $W_0\sim 0.1~\mathrm{mN/m}$, $\lambda\sim 6~\mathrm{nm}$), we get $N_k\sim 1$ and $N_K\sim 6\cdot10^{-3}$, in agreement with the fact that they have no spring instabilities and a little effect of the surface deformations on the structural force profile.
\item For a previous AFM study with polyelectrolytes~\cite{Moazzami-Gudarzi2016a} ($R\sim 2~\mathrm{\mu m}$, $k\sim 0.3~\mathrm{N/m}$, $K\sim 50~\mathrm{GPa}$, $W_0\sim 0.02~\mathrm{mN/m}$, $\lambda\sim 50~\mathrm{nm}$), we get $N_k\sim 10^{-1}$ and $N_K\sim 10^{-6}$. Also for a previous AFM study with colloidal suspensions~\cite{Schon2018a} ($R\sim 7~\mathrm{\mu m}$, $k\sim 0.03~\mathrm{N/m}$, $K\sim 50~\mathrm{GPa}$, $W_0\sim 0.005~\mathrm{mN/m}$, $\lambda\sim 70~\mathrm{nm}$), we get $N_k\sim 7\cdot 10^{-1}$ and $N_K\sim 4\cdot 10^{-7}$. This is in agreement with the fact that both studies have no spring instabilities and a little effect of the surface deformations on the structural force profile.
\end{itemize}

\noindent If the surface deformations have a little effect on the structural force profile ($N_K<1$, i.e. solids compression negligible compared to liquid compression), a fit with a semi-empirical exponentially decaying harmonic oscillation (equation~\ref{eq:structural_force_exp_cos_fit}) or a variation from it can be attempted. On the contrary, if the surface deformations have a strong effect on the structural force profile ($N_K>1$, i.e. solids compression dominant compared to liquid compression), a fit with our heuristic formulation (equation~\ref{eq:structural_force_DMT_fit}) or a variation from it can be attempted. For an intermediate situation where the surface deformations have a moderate effect on the structural force profile ($N_K \sim 1$, i.e. solids compression of the same order than liquid compression), more sophisticated models are required, like the energy minimization approach proposed by Hoth et al.~\cite{Hoth2014a}.

\section{Conclusions}
\label{sec:Conclusions}

By simultaneously measuring forces and characterizing \textit{in-situ} the geometry of the contact, we have shown that the mechanical deformations of the confining solids can have a strong influence on surface force measurements. Although this paper focus on the analysis of specific SFB experiments with dry atmosphere and an ionic liquid, we think some of the conclusions listed below may be of general interest for the whole community of surface force measurements.

\begin{itemize}
\item SFA experiments are not always in the JKR regime but can be in the DMT regime, typically for situations of moderate adhesion over a range of a few nanometers. Using the correct model of contact mechanics is crucial, notably for quantitative investigations of adhesion or friction. \textcolor{black}{The two regimes of contact are usually distinguished by calculating the value of the Maugis parameter from an estimate of the range of the attractive forces; it is in fact more accurate to look at the contact radius before jump-out.}
\item In classical SFA experiments using mica sheets glued on glass lenses, the mica does not only bend but can also experiences a nanometric compression.
\item This compression has to be taken into account for a proper calibration of the undeformed mica thickness in dry atmosphere, for example by fitting the relation between the contact radius and the force with the JKR model. The usual procedure, that consists in taking the jump-in point as a reference, leads to an underestimation of the mica thickness and an equivalent outward shift of the force profile measured after injecting the liquid. The error is $\sim 1~\mathrm{nm}$ for a $\sim 7~\mathrm{\mu m}$-thick mica, and is expected to decrease with the mica thickness, albeit always present.
\item For any system showing a structural force profile with SFA or AFM, the exponentially decaying harmonic oscillation due to local variations of liquid density is \textit{a priori} convoluted with the mechanical response of the confining solids.
\item To interpret correctly the detailed shape of the structural force profile, it is necessary to know the degree of convolution, that can be estimated with a scaling. Compression in the solids is dominant over compression in the liquid typically for simple liquids (large energies, small length-scales) in the SFA (large radius of curvature). For SFA experiments with mica sheets glued on glass lenses, the influence of mica compression is more important and independent of the mica thickness at low loads; and smaller and reduced for thinner mica at large loads. This effect is expected to be even more important at all loads when mica is replaced by a softer layer (like EPON glue)~\cite{Britton2014a,vanEngers2017a,vanEngers2018a}, or at high loads if the distance measurement includes the indentation of the whole solid bodies (not only the top layers)~\cite{Garcia2017a}.
\item When the solids compression is dominant compared to liquid compression, a fit of the structural force profile with a semi-empirical exponentially decaying harmonic function is not appropriate. Heuristic formulations can be considered, based on extensions of contact mechanics models to situations where the solid surfaces confine a structured liquid film.
\end{itemize}



\section{Supplementary Material}
\label{sec:SupplementaryMaterial}

See supplementary material for a first figure showing the random, sub-nanometric, shifts of the force profiles observed from run to run, and a second figure comparing the measured structural force profile to an exponentially decaying harmonic oscillation.

\section*{Acknowledgments}
\label{sec:Acknowledgments}

S.P. and R.L. are supported by The Leverhulme Trust (RPG-2015-328) and the ERC (under Starting Grant No. 676861, LIQUISWITCH). R.L. is supported by the EPA Cephalosporin Junior Research Fellowship and Linacre College (University of Oxford).


\bibliography{Manuscript}

\end{document}